\documentclass[a4paper]{spie}  

\usepackage{enumitem}
\usepackage{amsmath,amsfonts,amssymb}
\usepackage{graphicx}
\usepackage[colorlinks=true, allcolors=blue]{hyperref}
\usepackage{bbm}
\usepackage{caption}
\usepackage[skip=0.333\baselineskip]{subcaption}
\usepackage{floatrow}
\usepackage{multirow}
\newfloatcommand{capbtabbox}{table}[][\FBwidth]

\title{Learning Physics-Inspired Regularization for Medical Image Registration with Hypernetworks}

\author[a,b]{Anna Reithmeir}
\author[a,b,c,d]{Julia A. Schnabel}
\author[a,c]{Veronika A. Zimmer}
\affil[a]{School of Computation, Information and Technology, Technical University of Munich, Munich, Germany}
\affil[b]{Munich Center for Machine Learning, Munich, Germany}
\affil[c]{Institute of Machine Learning in Biomedical Imaging, Helmholtz Munich, Neuherberg, Germany}
\affil[d]{School of Biomedical Engineering and Imaging Sciences, King’s College London, London, UK}

\begin{document} 
\maketitle

\begin{abstract}
Medical image registration aims to identify the spatial deformation between images of the same anatomical
region and is fundamental to image-based diagnostics and therapy. 
To date, the majority of the deep learning-based registration methods employ regularizers that enforce global spatial smoothness, e.g., the diffusion regularizer. 
However, such regularizers are not
tailored to the data and might not be capable of reflecting the complex underlying deformation. 
In contrast,
physics-inspired regularizers promote physically plausible deformations. One such regularizer is the linear elastic
regularizer, which models the deformation of elastic material. 
These regularizers are driven by parameters that
define the material’s physical properties. 
For biological tissue, a wide range of estimations of such parameters can
be found in the literature, and it remains an open challenge to identify suitable parameter values for successful
registration. 

To overcome this problem and to incorporate physical properties into learning-based registration,
we propose to use a hypernetwork that learns the effect of the physical parameters of a physics-inspired regularizer on the resulting spatial deformation field. In particular, we adapt the HyperMorph framework to learn
the effect of the two elasticity parameters of the linear elastic regularizer. 
Our approach enables the efficient
discovery of suitable, data-specific physical parameters at test time. To the best of our knowledge, we are the
first to use a hypernetwork to learn physics-inspired regularization for medical image registration. 
We evaluate
our approach on 3D intra-patient lung CT images. 
The results show that the linear elastic regularizer can
yield comparable results to the diffusion regularizer in unsupervised learning-based registration while predicting deformations with fewer foldings. With our method, the adaptation of the physical parameters to the data can successfully be performed at test time. Our code is available at \url{https://github.com/annareithmeir/elastic-regularization-hypermorph}.

\end{abstract}

\section{INTRODUCTION}
In recent years, deep learning has driven forward the field of medical image registration \cite{vm, lapirn, learn2reg2022, huang}. Once a registration network is trained, it can register a new image pair in only a few seconds. This is a crucial advantage to traditional registration algorithms since (near) real-time registration is often required in clinical practice, e.g., in radiotherapy and image-guided surgery \cite{Rueckert2011, TEUWEN2022330}.

For successful registration, regularization is essential as it can drive the registration process towards anatomically feasible solutions. 
The majority of the popular learning-based registration methods employ a regularizer to control the global smoothness of the deformation. Two frequent choices are the total variation and diffusion regularizers.
In fact, 14 out of the 21 best-performing algorithms in the 2022 Learn2Reg challenge \cite{learn2reg2022} and many state-of-the-art methods such as VoxelMorph \cite{vm} and LapIRN \cite{lapirn} use the standard diffusion regularizer. 
Since these regularizers impose simplified constraints on the deformation field, they might not be capable of reflecting the complex motion of the underlying anatomical structures. 

\begin{figure}[t]
    \centering
    \includegraphics[width=0.65\textwidth]{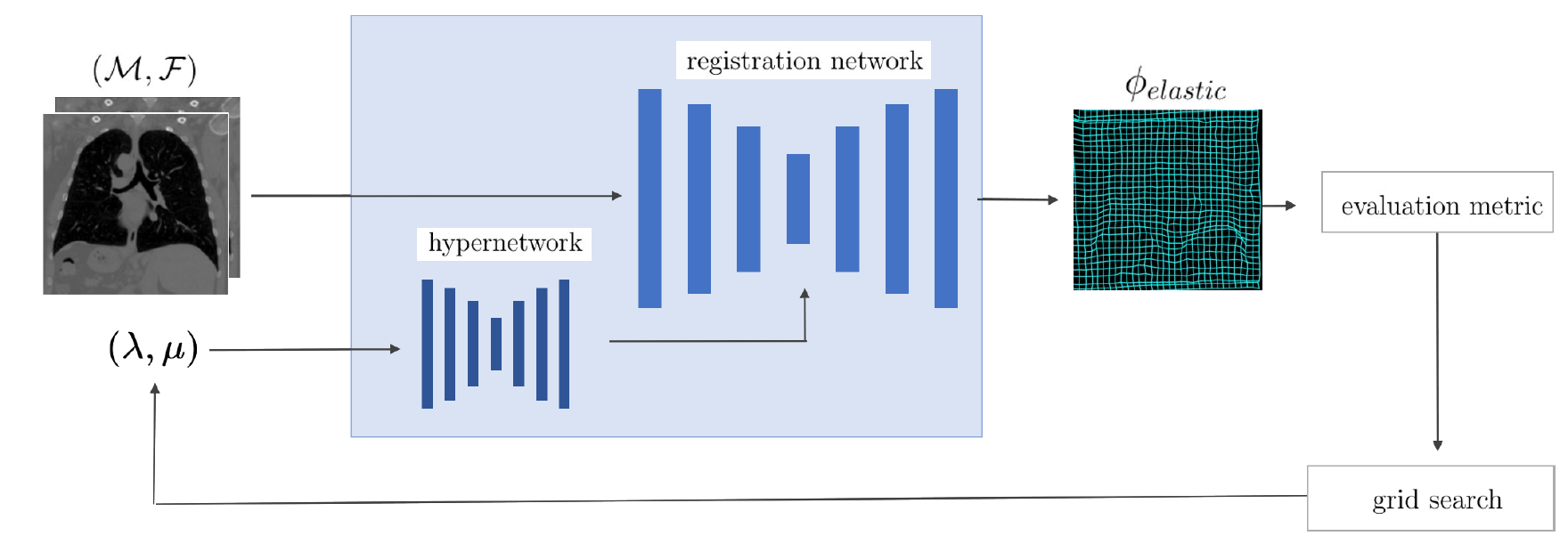}
    \caption{Overview of our approach. During training, a hypernetwork learns the effect of the physical parameters $\lambda, \mu$ of the linear elastic regularizer on the deformation field. At test time, the trained network takes the physical parameters as inputs and generates a deformation field that follows the physical properties specified by the parameter values. With a grid search, the optimal parameter values can then be identified according to a desired heuristic.}
    \label{fig:overview}
\end{figure}
 
Apart from spatial smoothness, regularization can promote the physical plausibility of the deformation. This is often desired in the medical context, for example, in intra-patient registration where images are registered that are taken of the same subject at different points in time.
Physical plausibility can be obtained with the help of physics-inspired regularizers, which model the physical properties of real-world material. One such regularizer is the linear elastic regularizer \cite{broit1981}, which models the behavior of elastic material under the application of external forces. 
Nevertheless, the practical application of physics-inspired regularizers is hindered by the challenge of identifying suitable values for the parameters that define the physical properties and thus drive these regularizers. For example, the linear elastic regularizer is driven by two parameters that define the elasticity properties of the material. While elasticity parameters have been determined for many types of material, they are ambiguous for living biological tissue. Various elasticity parameter estimations for specific tissue types can be found in the literature. However, they can vary not only in several orders of magnitude but also in their ratio \cite{hagemann}. Table  \ref{tab:estimations} shows an exemplary selection of brain and lung tissue estimations. 

The frequent presence of pathologies in medical images and patient-specific physical properties pose further difficulties. Previous works indicate that elasticity properties can differ between healthy and non-healthy subjects. For example, the lung tissue of patients with fibrotic lung disease is estimated to be stiffer than that of healthy subjects \cite{patient-specific}, and cancerous tissue can be stiffer than healthy tissue \cite{canceroustissue}. Apart from pathologies, the patient's age has an influence on bio-mechanical properties. For example, the stiffness of arteries increases with age while that of the skin decreases \cite{age-specific}.

Due to the challenges above, the a priori selection of physical parameter values for image registration is not straightforward and the parameters need to be adjusted not only to the anatomies present in the images but also to each specific image pair. As a result, physics-inspired regularizers are rarely applied in deep learning-based registration methods. Notable exceptions are the works of Qin et al. \cite{qin2023}, Hu et al. \cite{hu2018}, and Min et al. \cite{elasticdlpinn}. Here, the physical parameters are either randomly sampled or based on estimations from the literature. The only work that applies the linear elastic regularizer in a learning-based registration method is the recently proposed physics-informed neural network (PINN) by Min et al. \cite{elasticdlpinn}, which learns the registration between point clouds that have been derived from medical images. 

Altogether, this drives the need for data-driven physical parameter estimation in the context of medical image registration.
To overcome the challenges presented above and to drive forward the use of physics-inspired regularization in learning-based image registration, we propose to employ a novel method that simultaneously registers image pairs and estimates appropriate elasticity parameters that are tailored toward the data. We employ a hypernetwork to learn the effect of the physical parameters of a physics-inspired regularizer on the resulting deformation field. Once the model is trained, the appropriate parameters can be discovered efficiently. 
An overview of our method can be seen in Figure \ref{fig:overview}. 

\begin{table}[t]
    \centering
    \begin{tabular}{lcccc}
    \hline\hline
        Anatomy & Source & $\lambda$ & $\mu$ & Ratio ($\lambda/\mu$)\\
        \hline
        Brain&Hartmann et al. \cite{hartmann} & 12483.3 &25  & 499.322\\
        &Kumaresan et al. \cite{kumaresan} & 540.8 & 22.5 & 24.035\\
        \hline
        Lung&Lai-Fook et al. \cite{laifook} & 45.33 &8 & 5.667\\
        &Brock et al. \cite{brock} & 15.51 & 1.72 & 9.017\\
        \hline\hline
    \end{tabular}
    \caption{Elasticity parameter estimations ($\mu,\lambda$) of living biological tissue found in the literature can vary in their absolute value and ratio. Further estimations can be found in Hagemann et al. \cite{hagemann}.}
    \label{tab:estimations}
\end{table}

The contributions of this work are as follows:
\begin{enumerate}[noitemsep,nolistsep]
    \item We propose to learn the effect of the physical parameters of a physics-inspired regularizer on the deformation field with a hypernetwork. In particular, we build upon the HyperMorph framework \cite{hypermorph} and focus on the elasticity parameters of the linear elastic regularizer \cite{broit1981}.
    \item Our method allows efficient discovery of data-specific physical parameters at inference time, replacing the difficult and ambiguous a priori choice of physical parameter values.
    \item We include, for the first time, the linear elastic regularizer in an unsupervised learning-based image registration framework. To the best of our knowledge, the only other learning-based registration method \cite{elasticdlpinn} that applies the linear elastic regularizer works with point cloud data. 
\end{enumerate}

\section{BACKGROUND}

\subsection{Image Registration}
Medical image registration aims to identify the spatial deformation between two or more images of the same anatomical region. These can be, for example, images from different subjects (inter-subject registration) or images from the same subject acquired at different points in time (intra-subject registration).
Typically, registration is formulated as an optimization problem that recovers the underlying spatial deformation $\phi:\mathbbm{R}^D\rightarrow\mathbbm{R}^D$ between a moving and a fixed image $M, F: \Omega \subset \mathbbm{R}^{D}\rightarrow\mathbbm{R}$ by the minimization of the following objective functional:
\begin{equation}
    \label{eq:opt}
    \mathcal{L}(F, M, \phi) = \mathcal{L}_{sim}(F, M\circ\phi
    ) +\alpha \mathcal{L}_{Reg}(\phi).
\end{equation}
The data similarity term $\mathcal{L}_{Sim}$ measures how different the two images are to each other, for example, with the mean squared error of the pixel-wise intensities or - in the multi-modal case - with mutual information.
The optimization problem in Eq. \ref{eq:opt} is ill-posed, and for a given image pair, multiple deformations can be found that map the images onto each other. However, not all such deformations are realistic. To restrict the solution space and to drive the optimization towards a solution that fulfills desired properties, a regularization term $\mathcal{L}_{Reg}$ is used. A frequently used regularizer is the diffusion regularizer 
\begin{equation}
\label{eq:diff}
    \mathcal{L}_{Reg}=||\nabla \phi||^2
\end{equation}
where $\nabla$
is the Laplace operator. The diffusion regularizer can promote the global smoothness of the deformation.
Other properties that can be included by regularization are, for example, diffeomorphisms\cite{beg}, inverse consistency\cite{invcons}, and volume preservation\cite{mansi}.
The regularization weight $\alpha\in\mathbbm{R}^+$ balances the data similarity term $\mathcal{L}_{Sim}$ and the regularizer $\mathcal{L}_{Reg}$. In the following, $\phi$ is parameterized as a dense displacement field $\mathbf{u}\in\mathbbm{R}^{m\times n\times k\times D}$ that describes the displacement for each pixel location $x$ with $\phi(x)=x+\mathbf{u}(x)$.

For learning-based registration, the energy functional in Eq. \ref{eq:opt} can be used for the unsupervised training of a neural network, as, e.g., in Voxelmorph \cite{vm} and LapIRN \cite{mok}. Typically, a registration network takes a pair of images as the input and predicts a deformation field that registers one image onto the other. An advantage of registration networks is that, once they are trained, the registration of a novel image pair is performed efficiently in a single forward pass. In contrast to learning-based methods, conventional registration approaches optimize Eq. \ref{eq:opt} for each image pair individually.

\subsection{Linear Elastic Regularization}
Apart from the global smoothness of the deformation, physical plausibility can be desired in medical image registration.
One regularizer that is inspired by the laws of physics is the linear elastic regularizer brought forward by Broit\cite{broit1981}. It models the continuum mechanics of isotropic elastic material that deforms under the application of external forces. 
In the context of image registration, the moving image $M$ is regarded as an elastic object to which forces are applied at each pixel location such that it deforms until it matches the fixed image $F$. 
Originally, the regularizer was expressed with the Navier-Lam\'e differential equations and solved iteratively with finite elements methods. Later, a variational formulation of the linear elastic regularizer was formulated\cite{modersitzki}:
\begin{equation}
    \mathcal{L}_{Reg}(\phi;\lambda,\mu)= \int_\Omega  \frac{\mu}{4}\sum_{i,j=1}^{D} (\partial_{x_i}u_j + \partial_{x_j}u_i )^2 + \frac{\lambda}{2} (\mathrm{div\ } \mathbf{u})^2 d\mathbf{x}
\end{equation}
with $\mathrm{div}$ denoting the divergence and $\partial_{x_i}u_j$ denoting the j-th component of the spatial derivative of $\mathbf{u}$ w.r.t. dimension $i$. 

The elasticity properties of the material and, in particular, the stress-strain relationships are described by the two elasticity parameters $\lambda,\mu\in\mathbbm{R}^+$ (also called the Lam\'e parameters).
The shear modulus $\mu$ indicates the stiffness and describes the material's response to shear stress. $\lambda$ does not directly have a physical meaning but is connected to the bulk modulus, which describes the material's reaction to compression. 
They are closely related to Young's modulus, which relates the stretching along one direction with the material's internal forces, and the Poisson ratio, which describes the relation of the longitudinal stretching and the lateral shrinking of elastic material. In linear elasticity theory, the strain is assumed to be linearly proportional to the external forces.
Since $\lambda,\mu$ determine the elasticity properties of the material, they drive the deformation of $M$ and thus the overall registration process. 

As stated above, the identification of $\lambda,\mu$ remains a challenge in the context of medical image registration. In prior works that apply the linear elastic regularizer, they are either based on estimations found in the literature \cite{schenkenfelder, elasticdlpinn}, tuned as hyperparameters detached from their physical meaning \cite{Ronovsky2017ElasticIR} or set without further explanation \cite{ens, Yanovsky2008UnbiasedVR}. As can be seen in Tab. \ref{tab:estimations}, tissue-specific estimations in the literature can vary substantially, and it seems that there is no consensus about which values to use in the context of medical image registration.

\subsection{Hypernetworks}
A hypernetwork is a small neural network that outputs the weights of another, larger network\cite{ha2017hypernetworks}. Originally, Ha et al.\cite{ha2017hypernetworks} introduced hypernetworks to generate weights for each layer of a recurrent neural network based on a layer embedding vector. 
Both networks are trained end-to-end with back-propagation.
In previous work, hypernetworks have been used for the tuning of hyperparameters. Lorraine and Duvenaud \cite{lorraine2018stochastic} show that hypernetworks can successfully learn a mapping from hyperparameters to the network weights. With this approach, no re-training of the main network is necessary for tuning the hyperparameters.

Recently, HyperMorph\cite{hypermorph} was proposed for tuning the regularization weight $\alpha$ in the context of medical image registration. It employs a hypernetwork to learn the effect of $\alpha$ on the weights of a VoxelMorph\cite{vm} registration network. Thus, the effect of $\alpha$ on the predicted deformation is also learned. During training, the hyperparameter value is randomly sampled in each iteration from a prior distribution and then given as input to the hypernetwork.
The original framework uses the standard diffusion regularizer of Eq. \ref{eq:diff} and samples $\alpha$ from the uniform distribution $U(0,1)$. At inference time, a desired level of smoothness can be specified and passed to the network as additional input. 
Due to the near real-time registration with a trained VoxelMorph network, HyperMorph allows the efficient tuning of the regularization weight at test time, e.g., through user interaction or gradient-based optimization with frozen model weights.

\section{METHOD}
We propose to employ a hypernetwork to learn the physics-inspired regularization for a registration network (Fig. \ref{fig:overview}). In particular, we learn the effect of the two elasticity parameters $\lambda,\mu$ of the linear elastic regularizer on the deformation.
The parameter values are not chosen prior to the training. Instead, they are variable hyperparameters that form an additional input to the overall network: While the registration network takes an image pair as the input and predicts a deformation field, the hypernetwork takes the elasticity parameters as inputs and outputs the weights of the registration network.
During training, the two parameters are uniformly sampled from prior distributions $\lambda\sim U(0,1), \mu\sim U(0,1)$. Both the hypernetwork and the registration network are jointly optimized in a fully unsupervised manner.
We use the local normalized cross-correlation (NCC) with window size $9^D$ as the similarity metric and the variational formulation of the linear elastic regularizer: 
\begin{equation} 
\label{eq:elasopt1}
    \mathcal{L}(M,F,\phi, \lambda, \mu)= (1-\alpha) NCC(F,M\circ\phi) +\alpha\int_\Omega  \frac{\mu}{4}\sum_{i,j=1}^{D} (\partial_{x_i}u_j + \partial_{x_j}u_i )^2 + \frac{\lambda}{2} (\mathrm{div\ } \mathbf{u})^2\ d\mathbf{x}.
\end{equation} 
To reduce the number of hyperparameters, we absorb the regularization weight $\alpha$ in $\lambda$ and $\mu$. The overall training loss of our approach is
\begin{equation} 
\label{eq:elasopt2}
    \mathcal{L}(M,F,\phi, \lambda, \mu)= (1-\lambda_{\alpha}-\mu_{\alpha})NCC(F,M\circ\phi) +\int_\Omega  \frac{\mu_{\alpha}}{4}\sum_{i,j=1}^{D} (\partial_{x_i}u_j + \partial_{x_j}u_i )^2 + \frac{\lambda_{\alpha}}{2} (\mathrm{div\ } \mathbf{u})^2\ d\mathbf{x}.
\end{equation}  
To maintain a similar balance of the two loss terms as in Eq. \ref{eq:elasopt2}, we impose the constraint $0\leq\lambda_{\alpha} + \mu_{\alpha}\leq1$ on the two parameters.

At test time, the trained network takes an image pair $(M, F)$ and specified values $\lambda_{\alpha}, \mu_{\alpha}$ as inputs. It then predicts the deformation 
which is regularized according to the given elasticity properties. Since the elasticity parameters are variable inputs of the network, with our method, the parameter space of $\lambda_{\alpha},\mu_{\alpha}$ can be efficiently explored at test time: For any combination of parameter values, the corresponding regularized deformation is predicted by a forward pass through the network.
Thus, suitable elasticity parameter values can be identified without the need to re-train the network: The parameter combination that obtains the best registration result according to a desired heuristic can be found, for example, with a discrete grid search over the parameter space. 
We build upon the HyperMorph framework\cite{hypermorph} and use the default backbone architecture, including a VoxelMorph registration network\cite{vm}. For simplicity, we refer to the two parameters $\lambda_{\alpha},\mu_{\alpha}$ as $\lambda,\mu$ in the rest of this work.

\section{EXPERIMENTAL EVALUATION}
\subsection{Implementation and Datasets}
We implement our method in TensorFlow 2.12.1 snd use the Adam optimizer \cite{kingma2014adam}, a learning rate of 1e-4 and 250 training epochs for all experiments. The elastic regularizer is implemented with forward finite differences. All experiments are carried out on a machine with a NVIDIA RTXA6000 GPU. 

Two open-source lung CT datasets are used for the evaluation of our method: (1) The lung CT dataset\cite{lungctdata} (in the following denoted as L2R-lungCT)  and (2) the NLST dataset\cite{nlst} from the Learn2Reg challenge. For the first dataset, we use a train/validation/test split of 20/4/6. This dataset comprises intra-patient inhale and exhale images of size $192\times192\times208$ that are affinely pre-aligned. We resample the images to isotropic pixel spacing of 2mm. For the second dataset, we use a train/validation/test split of 170/10/30. The images are of size $224\times 192\times 224$ and have a pixel spacing of 1.5mm.
All images are clipped to the range (-1100,1518) and normalized prior to the training. Our code is available at \url{https://github.com/annareithmeir/elastic-regularization-hypermorph}.

For the evaluation, we use the Dice score (DSC) between the deformed moving and the fixed segmentation volume, as well as the target registration error (TRE) between the deformed moving and fixed keypoints. Furthermore, we assess the fraction of negative Jacobian determinant values ($\%$negJacDet) of the predicted deformation field which indicates the amount of foldings in the deformation. Since in the L2R-lungCT dataset, the fixed images are artificially cropped such that the lung segmentation is only partly visible, for this dataset, we perform the evaluation only inside the image domain of the cropped fixed image. For the NLST dataset, we evaluate the whole volume.

\subsection{Results}
\subsubsection{Linear Elastic Regularization with Fixed Elasticity Parameters} 
\begin{figure}
    \centering
    \includegraphics[width=\textwidth]{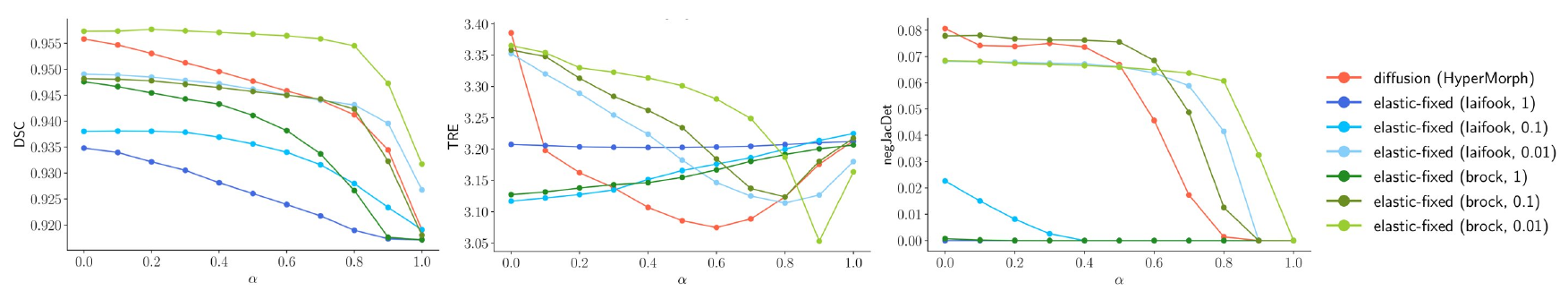}
    \caption{Mean DSC, TRE, and $\%$negJacDet for the L2R-lungCT test set with the HyperMorph model and different regularizer settings. For the linear elastic regularizer, the elasticity parameters are fixed to the values of Tab. \ref{tab:estimations} and weighted with the factor indicated in the legend. The input of the hypernetwork is the regularization weight $\alpha$ in all cases.}
    \label{fig:experiment1}
\end{figure}

In the first set of experiments, we investigate how the linear elastic regularizer performs compared to the diffusion regularizer with the standard HyperMorph framework. In all cases, the input to the hypernetwork is the regularization weight $\alpha$. We use the training loss of Eq. \ref{eq:elasopt1} and fix the elasticity parameters $\lambda,\mu$ of the linear elastic regularizer to values found in the literature, taken from Lai-Fook et al.\cite{laifook} ($\lambda=45.33,\mu=8$) and Brock et al.\cite{brock} ($\lambda=15.51,\mu=1.72$). 
Fig. \ref{fig:experiment1} shows the evaluation metrics on the test set for different values of $\alpha$. 

As expected, the Dice scores decrease for higher $\alpha$ since the deformation is more constrained if more weight is given to the regularization. 
Visual inspection of the results indicates that the proposed values result in very small deformations also for low $\alpha$. With the elasticity values above, the regularizer constrains the registration too much. 
Thus, we scale down the suggested values of $\mu,\lambda$ by a factor of $0.1$ and $0.01$ while keeping their ratio. It can be seen that lower elasticity parameter values result in higher DSC and that with the down-scaled elasticity parameters, a high DSC is maintained for a large range of $\alpha$. In comparison, the DSC with the diffusion regularizer drops consistently with increasing $\alpha$. In one case (brock*0.01 with $\lambda=0.1551,\mu=0.0172$), the linear elastic regularizer outperforms the diffusion regularizer for all $\alpha$.

Furthermore, it can be seen that with higher $\alpha$, the fraction of negative Jacobian determinants (\%negJacDet) decreases. This indicates that the more regularization is applied, the fewer foldings are present in the predicted deformation. With the original elasticity parameters, no folding is present due to the small magnitude of the deformation. For the down-scaled parameters, a comparable amount of foldings is obtained as with the diffusion regularizer for low $\alpha$. With the latter, the \%negJacDet drops from around $\alpha>0.5$ while this occurs only with higher $\alpha$ for the down-scaled elasticity parameters. 
Furthermore, for low $\lambda,\mu$, the TRE decreases with higher $\alpha$ while increasing again for $\alpha=1$, which is the highest possible regularization level. While with the diffusion regularizer, the TRE decreases already with small alpha and increases again for $\alpha>0.6$, with the linear elastic regularizer, the TRE decreases at a lower rate but only increases for $\alpha\geq 0.8$. For $\alpha>0.8$, the down-scaled elasticity parameters achieve a generally lower TRE than the diffusion regularizer. 

The lowest TRE is achieved with the linear elastic regularizer, the parameter values of brock*0.01, and $\alpha=0.9$. This parameter configuration also achieves the highest DSC at that level of regularization.
In general, the values from Lai-Fook et al.\cite{laifook} perform worse than the ones from Brock et al.\cite{brock}.
Visual inspection of the predicted deformations shows that with the diffusion regularizer and $\alpha\geq0.6$ as well as with the linear elastic regularizer and $\alpha\geq 0.8$, the cropping of the fixed images leads to severe foldings outside of the fixed image domain. This is due to the fact that the regions that are visible in the moving but not in the fixed images get pushed into the fixed image domain. This could be a possible reason for the delayed decrease in TRE and \%negJacDet with the linear elastic regularizer compared to the diffusion regularizer. To validate this hypothesis, measures that constrain the deformation outside of the fixed image domain should be explored in the future.

This first set of experiments shows that with small elasticity parameters, the linear elastic regularizer yields similar results to the diffusion regularizer
and that the registration performance highly depends on the specific values of $\lambda,\mu$. Resorting to elasticity estimations from the literature does not necessarily lead to good registration results. Thus, a method that identifies optimal values for $\lambda,\mu$ is needed for successful physics-inspired regularization with the linear elastic regularizer.

\subsubsection{Learning Linear Elastic Regularization with a Hypernetwork}
Next, we evaluate our proposed method on the L2R-lungCT and NLST datasets. 
In contrast to the first experiment, the two elasticity parameters $\lambda,\mu$ are variable hyperparameters that form the input of the hypernetwork.
We train the model with the training loss of Eq. \ref{eq:elasopt2}.
The main advantage of our method is that it offers the possibility to efficiently explore the elasticity parameter space at test time: Registration can be performed with different hyperparameter combinations by multiple forward passes through the trained network. The predicted registration results can then be evaluated for different elasticity parameter configurations, and the optimal parameter combination can be selected according to a desired heuristic. 
\begin{figure}[t]
    \centering
    \begin{subfigure}[b]{0.32\textwidth}
        \centering
         \includegraphics[width=0.82\textwidth]{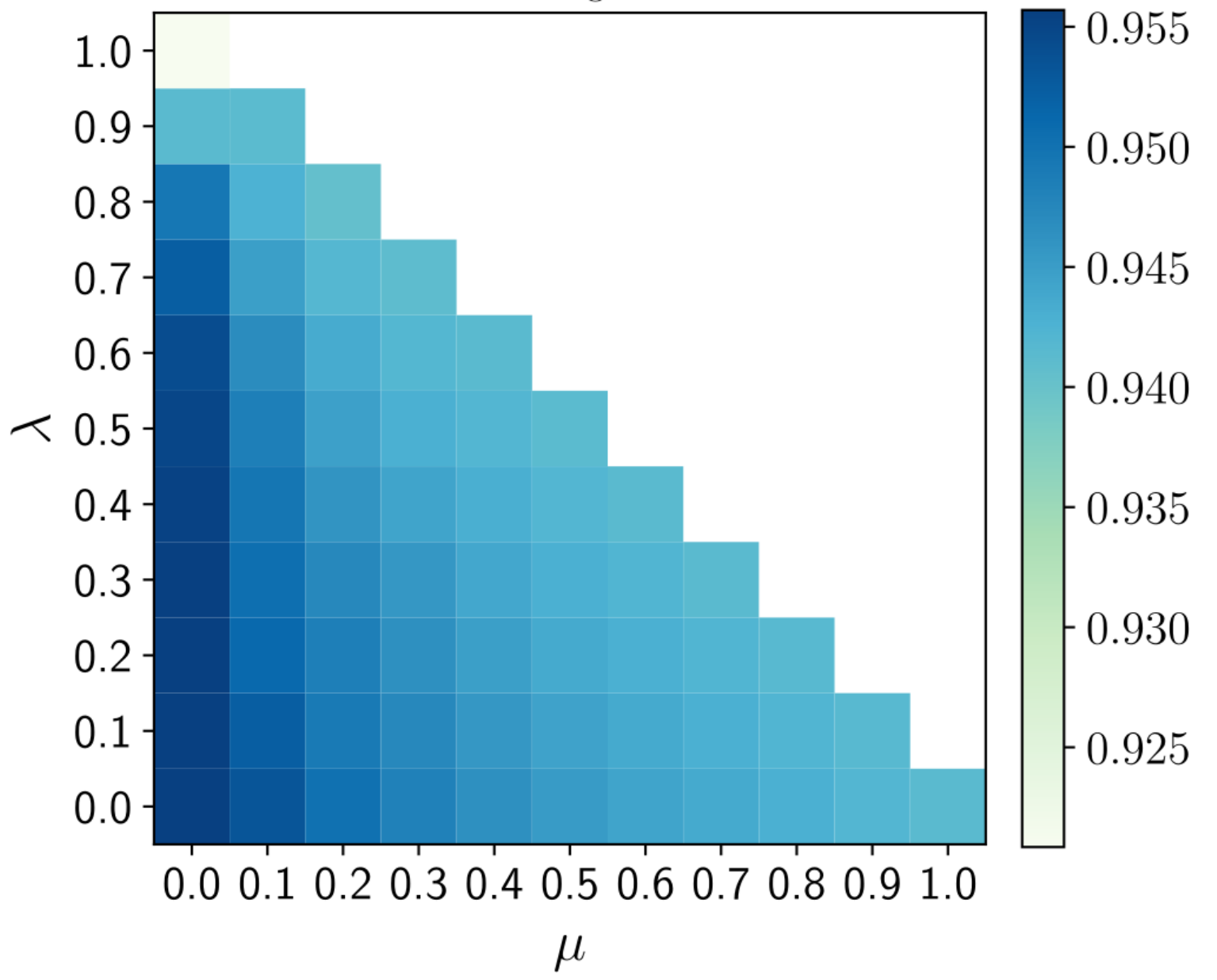}
         \caption{DSC}
         \label{fig:experiment2lungdsc}
     \end{subfigure}
     \begin{subfigure}[b]{0.32\textwidth}
        \centering
         \includegraphics[width=0.9\textwidth]{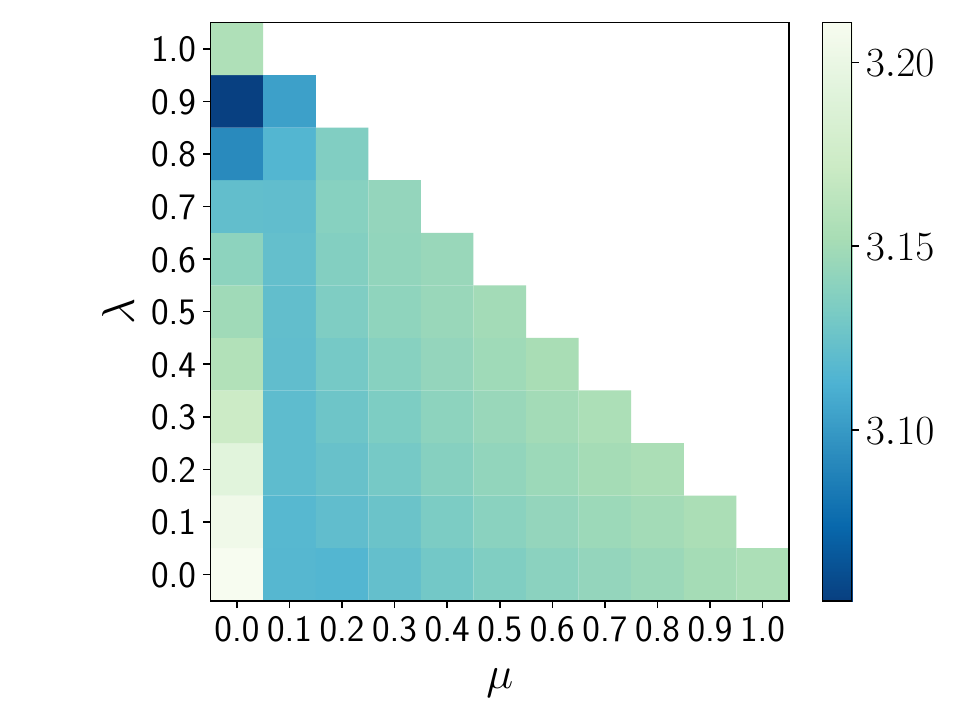}
         \caption{TRE}
         \label{fig:experiment2lungtre}
     \end{subfigure}
     \begin{subfigure}[b]{0.32\textwidth}
        \centering
         \includegraphics[width=0.9\textwidth]{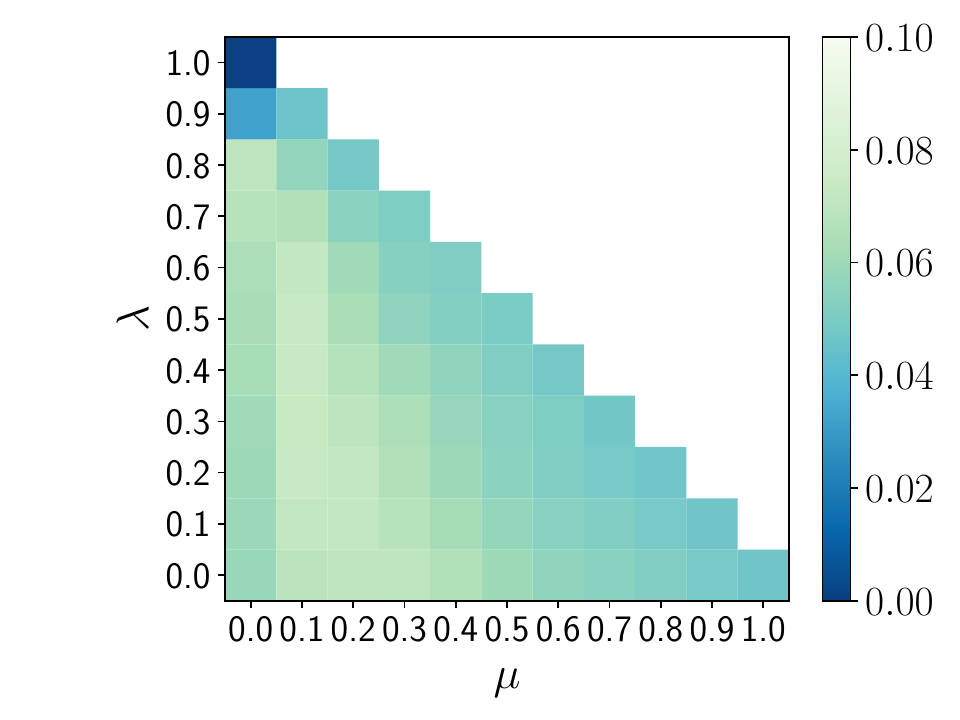}
         \caption{$\%$negJacDet}
         \label{fig:experiment2lungnegjac}
     \end{subfigure}\\
     \begin{subfigure}[b]{0.32\textwidth}
        \centering
         \includegraphics[width=0.82\textwidth]{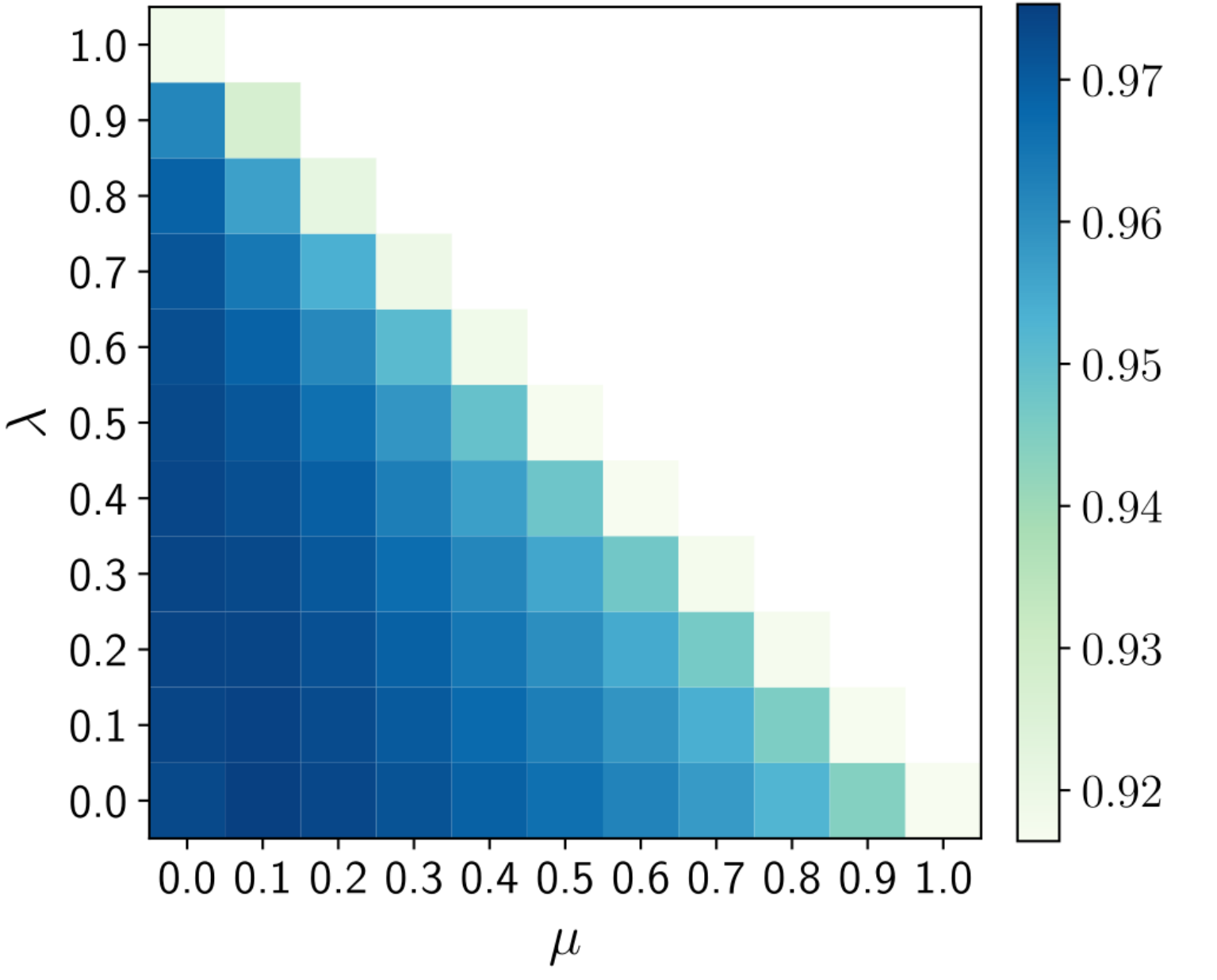}
         \caption{DSC}
         \label{fig:experiment2nlstdsc}
     \end{subfigure}
     \begin{subfigure}[b]{0.32\textwidth}
        \centering
         \includegraphics[width=0.9\textwidth]{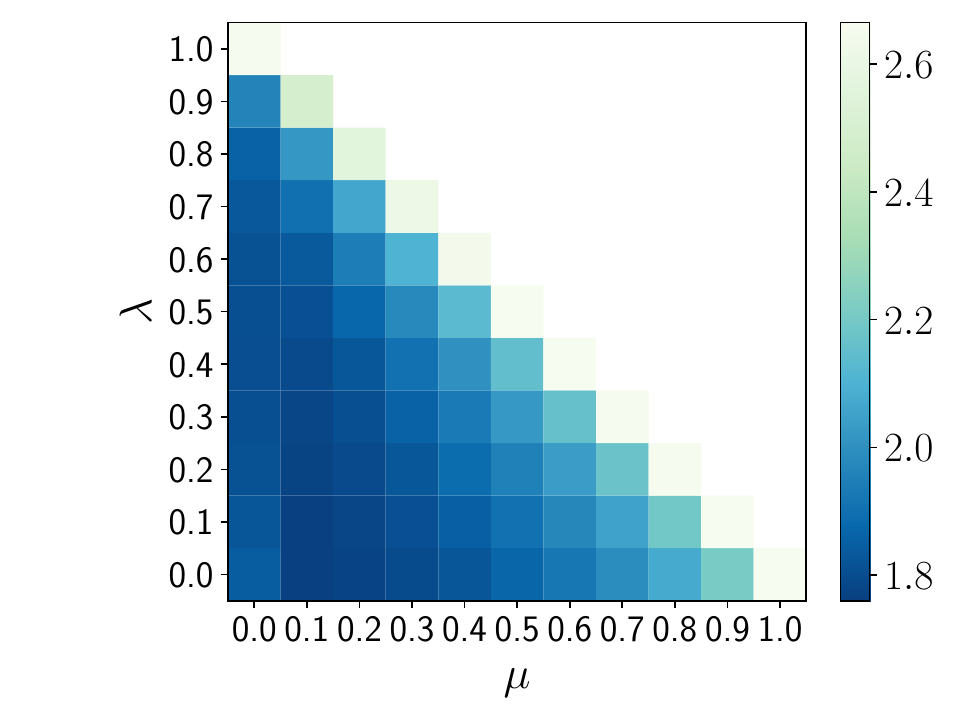}
         \caption{TRE}
         \label{fig:experiment2nlsttre}
     \end{subfigure}
     \begin{subfigure}[b]{0.32\textwidth}
        \centering
         \includegraphics[width=0.9\textwidth]{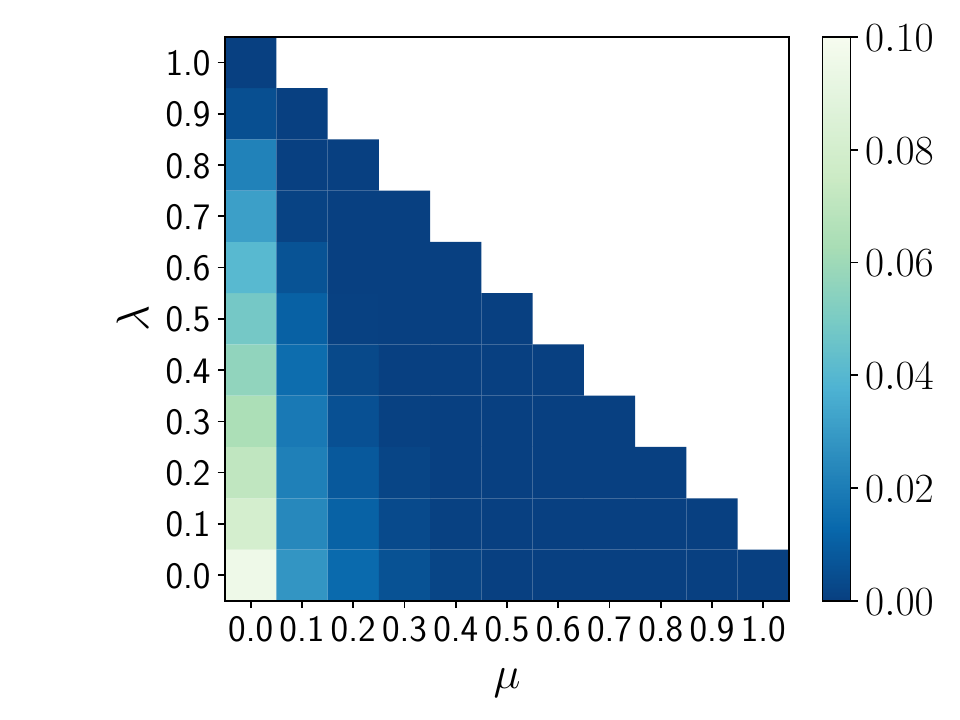}
         \caption{$\%$negJacDet}
         \label{fig:experiment2nlstnegjac}
     \end{subfigure}
    \caption{Mean DSC, TRE, and \%negJacDet obtained with the test set of the L2R-lungCT (top row, Figs. \ref{fig:experiment2lungdsc},\ref{fig:experiment2lungtre} and \ref{fig:experiment2lungnegjac}) and NLST (bottom row, Figs. \ref{fig:experiment2nlstdsc},\ref{fig:experiment2nlsttre} and \ref{fig:experiment2nlstnegjac}) datasets for different combinations of the elasticity parameters $\lambda, \mu$ . The influence of the hyperparameter values on the evaluation metrics varies visibly between the two datasets.}
    \label{fig:exp2both}
\end{figure}
To explore the parameter space, we perform a discrete grid search over $\lambda,\mu$ with a grid resolution of $0.1$.
The mean evaluation metrics on the two test sets are shown in Fig. \ref{fig:exp2both} for different combinations of parameter values. 
Note that due to the constraint $0\leq\lambda_{\alpha} + \mu_{\alpha}\leq1$ the upper right triangle of the combinations is not feasible. 

It can be seen that the influence of the elasticity parameter values on the results differs between the two datasets. 
While for the L2R-lungCT dataset, the DSC shows high sensitivity to $\mu$ and high DSC are only achieved for very low $\mu$, this is not the case for the NLST dataset (compare Fig. \ref{fig:experiment2lungdsc} and Fig. \ref{fig:experiment2nlstdsc}). Here, the DSC is similarly sensitive to changes in both parameters. 
For $\mu=0$, with the L2R-lungCT dataset, an increase in DSC compared to $\mu=0.1$ is observed, while for the NLST dataset, these values for $\mu$ lead to comparable DSC. 
Furthermore, the marginal parameter combinations where $\lambda_{\alpha} + \mu_{\alpha}=1$ perform similarly to neighboring combinations for the L2R-lungCT dataset. For the NLST dataset, however, a substantial drop in performance can be observed for the marginal combinations.

Moreover, in both datasets a low \%negJacDet is achieved for high $\mu$ (L2R-lungCT: $\mu>0.6$, NLST: $\mu>0.4$) where no sensitivity to $\lambda$ is observed (compare Fig. \ref{fig:experiment2lungnegjac} and Fig. \ref{fig:experiment2nlstnegjac}). The \%negJacDet varies with changing $\lambda$ for lower values of $\mu$, and when $\mu=0$ a high sensitivity to $\lambda$ is observed.
Furthermore, a low TRE is achieved only for low $\mu$ ($0.1 <\mu<0.3$) (compare Fig. \ref{fig:experiment2lungtre} and \ref{fig:experiment2nlsttre}). 
All in all, the results reveal a high sensitivity of all three evaluation metrics to $\mu$ and that the influence of the elasticity parameters on the registration results varies between different datasets. 

\begin{table}[t]
    \centering
    \begin{tabular}{ccc|cccc}
        \hline\hline
        Heuristic & dataset & model & hyperparameters & DSC & TRE & negFrac \\
        \hline
        \multirow{ 4}{*}{DSC} & LungCT & diffusion\cite{hypermorph} &$\alpha=0.1$&$0.955\pm 0.010$&$3.198\pm 0.569$&$0.074\pm 0.014$\\
        &LungCT & linear elastic (ours) &$\lambda=0.2,\mu=0.0$&\textbf{$0.956 \pm 0.012$}&\textbf{$3.192 \pm 0.678$}&\textbf{$0.059 \pm0.009$}\\
        &NLST &diffusion\cite{hypermorph}&$\alpha=0.2$&$0.977 \pm 0.014$&$1.749 \pm 0.593$&$0.060 \pm 0.023$\\
        &NLST &linear elastic (ours)&$\lambda=0.0,\mu=0.1$&$0.975 \pm0.017$&$1.759\pm0.619$&$0.028\pm0.013$\\
        \hline
        \multirow{ 4}{*}{TRE} &LungCT & diffusion\cite{hypermorph} &$\alpha=0.6$&$0.945\pm 0.012$&$3.075\pm 0.495$&$0.046\pm 0.013$\\
        &LungCT & linear elastic (ours) &$\lambda=0.9,\mu=0.0$&$0.941 \pm 0.012$&$3.103 \pm 0.401$&$0.046 \pm0.012$\\
        &NLST &diffusion\cite{hypermorph}&$\alpha=0.4$&$0.976 \pm 0.016$&$1.735 \pm 0.590$&$0.033 \pm 0.016$\\
        &NLST &linear elastic (ours)&$\lambda=0.0,\mu=0.1$&$0.975 \pm0.017$&$1.759\pm0.619$&$0.028\pm0.013$\\
        \hline\hline
    \end{tabular}
    \caption{At test time, the optimal parameter configuration can be identified according to a desired heuristic. We extract the parameters based on the mean DSC (top) and TRE (bottom) with the test set. For the proposed method, the elasticity values $\lambda,\mu$ are extracted. For the HyperMorph model with the diffusion regularizer, the regularization weight $\alpha$ is extracted. The proposed method yields comparable results to the hyperMorph model while predicting deformations with fewer foldings.}
    \label{tab:bestdsctre}
\end{table}
With the parameter space exploration, the optimal elasticity parameter combinations can be found with respect to a desired heuristic. To do so, we use the DSC and TRE as heuristics and extract the parameter combinations that optimize them individually. Similarly, the optimal $\alpha$ is identified for the diffusion regularizer of the original HyperMorph model. 
Table \ref{tab:bestdsctre} shows the evaluation metrics for the predicted registration that is obtained with the identified optimal parameters. It can be seen that the optimal parameter values can vary across the two heuristics and also across the datasets. However, with the linear elastic regularizer, a single parameter combination ($\lambda=0.0,\mu=0.1$) optimizes both heuristics for the NLST dataset. 

\begin{figure}[t]
    \centering
    \includegraphics[width=0.85\textwidth]{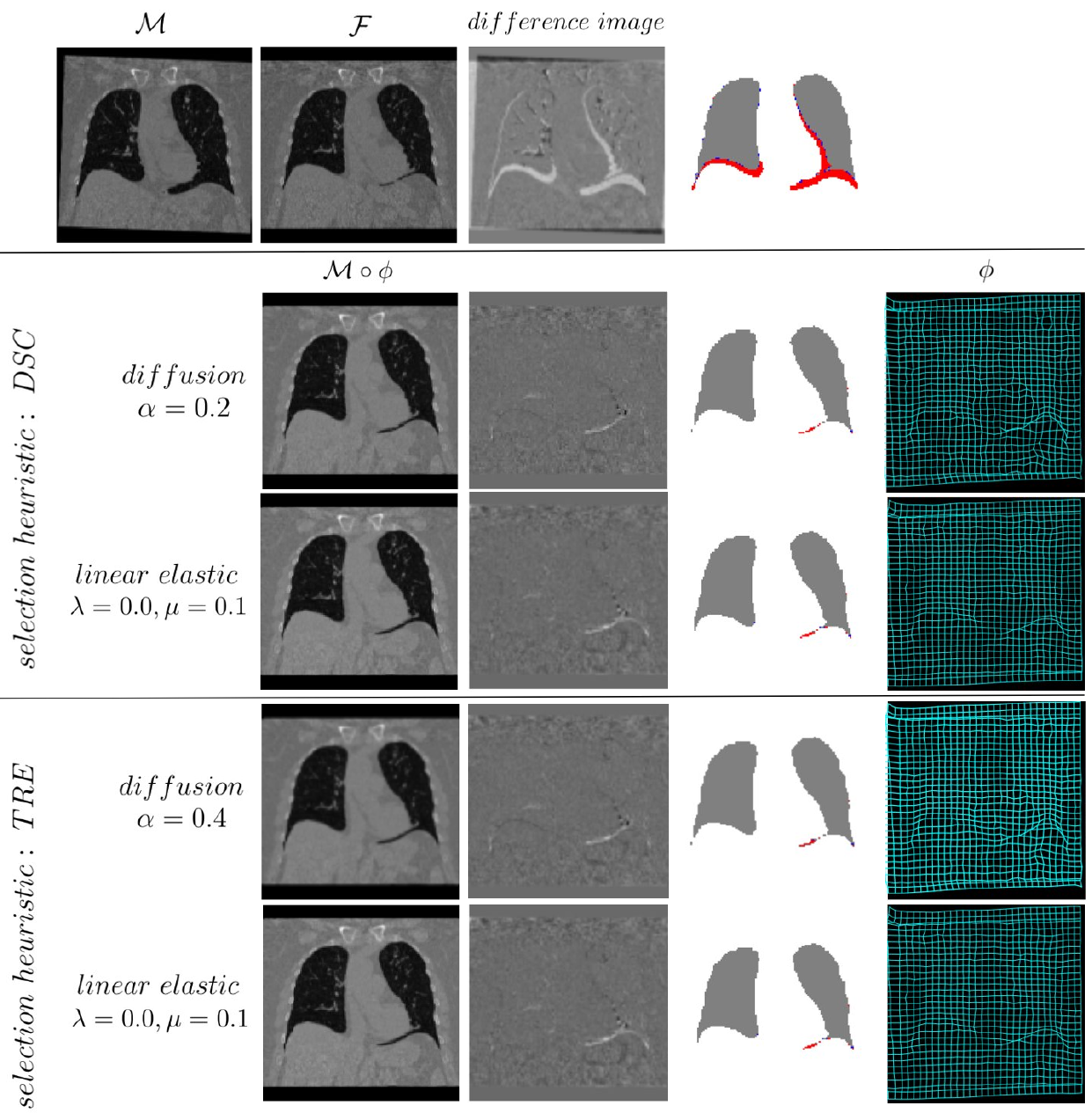}
    \caption{Registration results of a randomly chosen test sample from the NLST dataset for the identified optimal hyperparameter values (see Tab.\ref{tab:bestdsctre}). For each case, the predicted deformed moving image, the difference image between the fixed and the predicted segmentation map and between the fixed and the predicted image, and the deformation field are shown. The middle coronal slices are depicted.}
    \label{fig:plotall}
\end{figure}

Overall, the linear elastic regularizer yields comparable DSC and TRE to the diffusion regularizer while consistently showing fewer foldings in the predicted deformation (\%negJacDet). Visual inspection of the deformation fields reveals that the linear elastic regularizer constrains the deformation more than the diffusion regularizer, which can be seen by a generally smaller deformation magnitude. For small $\alpha$, the diffusion regularizer produces deformations with visible folding, while the linear elastic regularizer produces smoother deformations across all values of $\lambda,\mu$.
This observation is reflected in the qualitative registration results of Fig. \ref{fig:plotall}, which show the registration of a randomly chosen sample of the NLST dataset with the identified optimal parameters of Tab. \ref{tab:bestdsctre}.

The process of parameter space exploration and optimal value identification can easily be adapted to the specific needs of the user. First, more complex heuristics could be used for parameter optimization such as a linear combination of multiple evaluation metrics. Second, while in this work the exploration and parameter optimization is based on the whole test set, it could instead be performed on a specific image pair. Third, a multi-resolution grid search could be performed for fine-tuning.

\section{CONCLUSION AND OUTLOOK}
In this work, we proposed a novel unsupervised method to identify the elasticity parameter values of the linear elastic regularizer for medical image registration. Our method employs a hypernetwork which allows the efficient exploration of the elasticity parameter space at test time without the need to retrain the model. The evaluation of our method on two openly available lung CT datasets shows that the registration performance with the linear elastic regularizer depends on the chosen parameter values and that with our approach, suitable values can be identified at test time. Compared to the original HyperMorph model which uses the diffusion regularizer, our method yields comparable results in terms of the TRE and DSC with consistently fewer foldings in the predicted deformations while promoting physical plausibility. 

This work is a proof of concept and has several limitations. First, it employs a global regularizer. This might be suboptimal for multi-organ medical images since different anatomical structures have different elasticity properties. Second, with the absorption of the regularization weight into the elasticity parameters and the additional constraint on their sum, the direct physical interpretation of the identified parameters is lost. Third, the method is evaluated on lung CT images only. 
In the future, we plan to investigate how our method can be adapted to spatially varying regularization. 
Also, we aim to explore other physics-inspired regularizers such as the hyper-elastic regularizer \cite{hyperelastic} and to further evaluate the method on different anatomical structures such as the heart or the abdomen. 

\bibliography{report} 

\begin{thebibliography}{10}

\bibitem{vm}
Balakrishnan, G., Zhao, A., Sabuncu, M., Guttag, J., and Dalca, A., ``Voxelmorph: A learning framework for deformable medical image registration,'' {\em IEEE Transactions on Medical Imaging}~{\bf 38},  1788--1800 (2019).

\bibitem{lapirn}
Mok, T. C.~W. and Chung, A. C.~S., ``Large deformation diffeomorphic image registration with laplacian pyramid networks,'' {\em Medical Image Computing and Computer Assisted Intervention (MICCAI)} {\bf 12263},  211--221 (2020).

\bibitem{learn2reg2022}
Hering, A., Hansen, L., Mok, T. C.~W., et~al., ``{L}earn2{R}eg: Comprehensive multi-task medical image registration challenge, dataset and evaluation in the era of deep learning,'' {\em IEEE Transactions on Medical Imaging}~{\bf 42}(3),  697--712 (2023).

\bibitem{huang}
Huang, W., Yang, H., Liu, X., Li, C., Zhang, I., Wang, R., Zheng, H., and Wang, S., ``A coarse-to-fine deformable transformation framework for unsupervised multi-contrast {MR} image registration with dual consistency constraint,'' {\em IEEE Transactions on Medical Imaging}~{\bf 40}(10),  2589--2599 (2021).

\bibitem{Rueckert2011}
Rueckert, D. and Schnabel, J.~A.,  [{\em Medical Image Registration}{\nolinebreak\hspace{0.1em}]},  131--154, Springer, Berlin, Heidelberg (2011).

\bibitem{TEUWEN2022330}
Teuwen, J., Gouw, Z.~A., and Sonke, J.-J., ``Artificial intelligence for image registration in radiation oncology,'' {\em Seminars in Radiation Oncology}~{\bf 32}(4),  330--342 (2022).
\newblock Artificial Intelligence: Methods and Applications in Radiotherapy.

\bibitem{broit1981}
Broit, C.,  [{\em Optimal Registration of Deformed Images}{\nolinebreak\hspace{0.1em}]}, Graduate School of Arts and Sciences, University of Pennsylvania (1981).

\bibitem{hagemann}
Hagemann, A., Rohr, K., Stiehl, H., Spetzger, U., and Gilsbach, J., ``Biomechanical modeling of the human head for physically based, nonrigid image registration,'' {\em IEEE Transactions on Medical Imaging}~{\bf 18}(10),  875--884 (1999).

\bibitem{patient-specific}
Marinelli, J., Levin, D., Vassallo, R., Carter, R., Hubmayr, R., Ehman, R., and McGee, K., ``Quantitative assessment of lung stiffness in patients with interstitial lung disease using mr elastography: Quantifying lung stiffness in {ILD},'' {\em Journal of Magnetic Resonance Imaging}~{\bf 46} (01 2017).

\bibitem{canceroustissue}
Hoyt, K., Casta{\~n}eda, B., Zhang, M., Nigwekar, P., di~Sant'Agnese, P.~A., Joseph, J., Strang, J.~G., Rubens, D.~J., and Parker, K.~J., ``Tissue elasticity properties as biomarkers for prostate cancer.,'' {\em Cancer biomarkers : section A of Disease markers}~{\bf 4 4-5},  213--25 (2008).

\bibitem{age-specific}
Akhtar, R., Sherratt, M.~J., Cruickshank, J.~K., and Derby, B., ``Characterizing the elastic properties of tissues,'' {\em Materials Today}~{\bf 14}(3),  96--105 (2011).

\bibitem{qin2023}
Qin, C., Wang, S., Chen, C., Bai, W., and Rueckert, D., ``Generative myocardial motion tracking via latent space exploration with biomechanics-informed prior,'' {\em Medical Image Analysis}~{\bf 83},  102682 (2023).

\bibitem{hu2018}
Hu, Y., Gibson, E., Ghavami, N., Bonmati, E., Moore, C., Emberton, M., Vercauteren, T., Noble, J., and Barratt, D.~C., ``Adversarial deformation regularization for training image registration neural networks,'' (2018).

\bibitem{elasticdlpinn}
Min, Z., Baum, Z. M.~C., Saeed, S.~U., Emberton, M., Barratt, D.~C., Taylor, Z.~A., and Hu, Y., ``Non-rigid medical image registration using physics-informed neural networks,'' {\em Information Processing in Medical Imaging (IPMI)} {\bf 13939},  601--613 (2023).

\bibitem{hartmann}
Hartmann, U. and Kruggel, F., ``Erste klinische untersuchungen mit einem mechanischen finite-elemente-modell des menschlichen kopfes,'' in [{\em Bildverarbeitung f{\"u}r die Medizin 1998}{\nolinebreak\hspace{0.1em}]},  Lehmann, T., Metzler, V., Spitzer, K., and Tolxdorff, T., eds.,  59--63, Springer Berlin Heidelberg, Berlin, Heidelberg (1998).

\bibitem{kumaresan}
Kumaresan, S. and Radhakrishnan, S., ``Importance of partitioning membranes of the brain and the influence of the neck in head injury modelling,'' {\em Medical and Biological Engineering and Computing}~{\bf 34},  27--32 (2007).

\bibitem{laifook}
Lai-Fook, S.~J. and Hyatt, R.~E., ``Effects of age on elastic moduli of human lungs,'' {\em Journal of Applied Physiology}~{\bf 89}(1),  163--168 (2000).

\bibitem{brock}
Brock, K.~K., Sharpe, M.~B., Dawson, L.~A., Kim, S.~M., and Jaffray, D.~A., ``Accuracy of finite element model-based multi-organ deformable image registration,'' {\em Medical Physics}~{\bf 32},  1647--1659 (2005).

\bibitem{hypermorph}
Hoopes, A., Hoffmann, M., Fischl, B., Guttag, J., and Dalca, A.~V., ``Hypermorph: Amortized hyperparameter learning for image registration,'' {\em Information Processing in Medical Imaging (IPMI)} {\bf 12729},  3–17 (2021).

\bibitem{beg}
Beg, M.~F., Miller, M., Trouvé, A., and Younes, L., ``Computing large deformation metric mappings via geodesic flows of diffeomorphisms,'' {\em International Journal of Computer Vision}~{\bf 61},  139--157 (2005).

\bibitem{invcons}
Zhu, Y. and Lu, S., ``Swin-voxelmorph: A symmetric unsupervised learning model for deformable medical image registration using swin transformer,'' in [{\em Medical Image Computing and Computer Assisted Intervention -- (MICCAI)}{\nolinebreak\hspace{0.1em}]},  Wang, L., Dou, Q., Fletcher, P.~T., Speidel, S., and Li, S., eds.,  78--87, Springer Nature Switzerland, Cham (2022).

\bibitem{mansi}
Mansi, T., Pennec, X., Sermesant, M., Delingette, H., and Ayache, N., ``{LogDemons Revisited: Consistent Regularisation and Incompressibility Constraint for Soft Tissue Tracking in Medical Images},'' in [{\em {Proc. of Medical Image Computing and Computer-Assisted Intervention (MICCAI)}}{\nolinebreak\hspace{0.1em}]},  {\em LNCS} {\bf 6362},  652--659 (2010).

\bibitem{mok}
Mok, T. C.~W. and Chung, A. C.~S., ``Conditional deformable image registration with convolutional neural network,'' {\em Medical Image Computing and Computer Assisted Intervention (MICCAI)}~{\bf 12904},  35--45 (2021).

\bibitem{modersitzki}
Fischer, B. and Modersitzki, J., ``A unified approach to fast image registration and a new curvature based registration technique,'' {\em Linear Algebra and its Applications}~{\bf 380},  107--124 (2004).

\bibitem{schenkenfelder}
Schenkenfelder, B., Fenz, W., Thumfart, S., Ebenhofer, G., Stübl, G., Lumenta, D.~B., Reishofer, G., and Scharinger, J., ``Elastic registration of abdominal {MRI} scans and {RGB-D} images to improve surgical planning of breast reconstruction,'' in [{\em Annual Modeling and Simulation Conference (ANNSIM)}{\nolinebreak\hspace{0.1em}]},  (2021).

\bibitem{Ronovsky2017ElasticIR}
Ronovsky, A. and Vasatov{\'a}, A., ``Elastic image registration based on domain decomposition with mesh adaptation,'' {\em Advances in Electrical and Electronic Engineering}~{\bf 15},  322--330 (2017).

\bibitem{ens}
Ens, K., Schumacher, H., Franz, A., Fischer, B., Pluim, J., and Reinhardt, J., ``Improved elastic medical image registration using mutual information,'' in [{\em Medical Imaging: Image Processing}{\nolinebreak\hspace{0.1em}]},   {\bf 6512} (2007).

\bibitem{Yanovsky2008UnbiasedVR}
Yanovsky, I., Guyader, C.~L., Leow, A.~D., Toga, A.~W., Thompson, P.~M., and Vese, L.~A., ``Unbiased volumetric registration via nonlinear elastic regularization,'' (2008).

\bibitem{ha2017hypernetworks}
Ha, D., Dai, A.~M., and Le, Q.~V., ``Hypernetworks,'' {\em arXiv preprint arXiv:1609.09106}  (2017).

\bibitem{lorraine2018stochastic}
Lorraine, J. and Duvenaud, D., ``Stochastic hyperparameter optimization through hypernetworks,'' {\em arXiv preprint arXiv:1802.09419}  (2018).

\bibitem{kingma2014adam}
Kingma, D.~P. and Ba, J., ``Adam: A method for stochastic optimization,'' {\em arXiv preprint arXiv:1412.6980}  (2014).

\bibitem{lungctdata}
Hering, A., Murphy, K., and van Ginneken, B., ``{L}earn2{R}eg challenge: {CT} lung registration - training data [data set].,'' (2020).
\newblock Zenodo. http://doi.org/10.5281/zenodo.3835682.

\bibitem{nlst}
National Lung Screening Trial Research~Team, T. C. I.~A., ``Data from the national lung screening trial ({NLST}) [data set],'' (2013).

\bibitem{hyperelastic}
Burger, M., Modersitzki, J., and Ruthotto, L., ``A hyperelastic regularization energy for image registration,'' {\em SIAM Journal on Scientific Computing}~{\bf 35}(1),  B132--B148 (2013).

\end{thebibliography}
\bibliographystyle{spiebib} 

\end{document}